# Dynamic Risk Management in Cyber Physical Systems

Daniel Schneider, Jan Reich, Rasmus Adler and Peter Liggesmeyer

[1] Fraunhofer IESE, Fraunhofer-Platz 1, 67663 Kaiserslautern, Germany
`name.surname@iese.fraunhofer.de`

**Abstract.** Cyber Physical Systems (CPS) enable new kinds of applications as well as significant improvements of existing ones in numerous different application domains. A major trait of upcoming CPS is an increasing degree of automation up to the point of autonomy, as there is a huge potential for economic success as well as for ecologic and societal improvements. However, to unlock the full potential of such (cooperative and automated) CPS, we first need to overcome several significant engineering challenges, where safety assurance is a particularly important one. Unfortunately, established safety assurance methods and standards do not live up to this task, as they have been designed with closed and less complex systems in mind. This paper structures safety assurance challenges of cooperative automated CPS, provides an overview on our vision of dynamic risk management and describes already existing building blocks.

**Keywords:** Safety, Cyber-Physical Systems, Runtime Safety, Dynamic Risk Management.

## 1      Introduction

Over the last decades we witnessed a very strong trend of computerization and digitization. As a consequence, people's daily lives and ways of working have been undergoing significant transformations. For the future we are anticipating these trends to continue, maybe even further accelerate, and thus digital systems to be even more ubiquitous in everybody's daily lives. We expect them to permeate every facet of it, to support humans in any way imaginable, be it with respect to information, health, mobility or work. At the same time, work life will undergo further transformation as systems are increasingly assuming tasks that today are exclusively addressed by humans. Corresponding visions are being developed and pursued for many years now and a pretty impressive range of corresponding umbrella terms have been coined along the way, such as Ubiquitous Computing, Pervasive Computing, Ambient Intelligence, Internet of Things or Cyber-Physical Systems. In addition, there are domain specific umbrella terms such as Industry 4.0, Autonomous Driving, Connected Automated Mobility, Operating Room of the Future, Ambient Assisted Living, and so on. All these visions of future systems are essentially enabled by a trinity of technological characteristics – **automation, interconnection, and AI,** where automation is the end and interconnection and AI are the indispensable means.

The potential is huge. The embedded systems industry, for example, is a significant job-creator in Europe, contributing a 34% share to the world production of embedded systems with particular strengths in the automotive sector, aerospace and health [1]. In the Electronic Components and Systems Strategic Research and Innovation Agenda 2023 (ECS-SRIA), the EU puts specific emphasis on the SoS topic, but also use of AI and pushing automation feature very strongly [2]. Further, AI in general has been identified as strategic technology and AI strategies have been devised by the key economic powers around the globe including Germany [3], Europe [4], US [5] and China [6]. As we previously elaborated in [7], aside from creating opportunities to improve and optimize existing businesses, digitalization also supports new forms of business models. In that respect, data-driven technologies are one central diver, paving the path for data-driven business models [8]. The key modification from traditional business models is that data-driven business models exploit data as the key value proposition [9]. As dependable, collaborative and autonomous systems rely on advanced sensor technologies and create huge amounts of data, data-driven business models are strongly appealing. Another driver are completely new types of applications which are enabled due to automation. Take mobility as a service as an example, which might revolutionize the way mobility is realized in our societies.

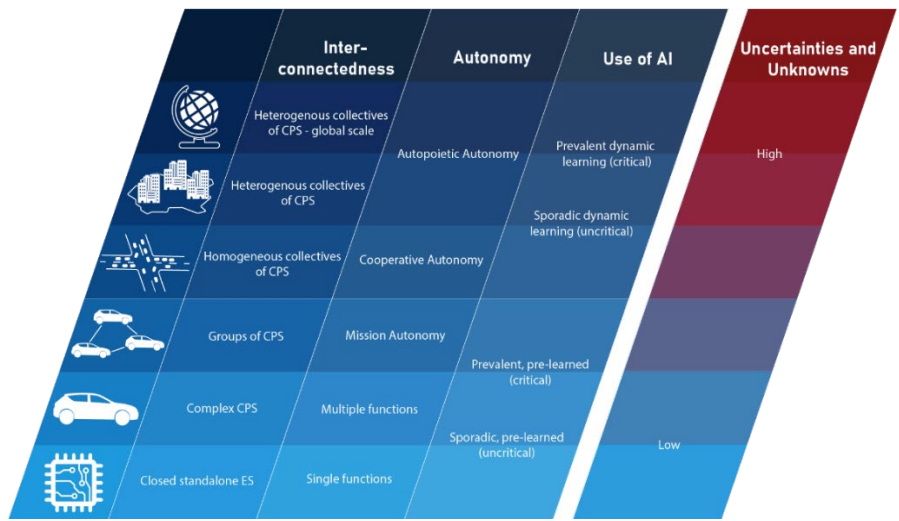

**Fig. 1.** Key trends as per [10]

But there are still significant challenges ahead which need to be tackled before this potential can be fully unlocked. These challenges can roughly be structured into three categories: *technological*, *structural/frame conditions* (e.g. laws, standards, interoperability, societal, education) and *assurance*. Technology seems to be relatively straight forward. Despite there always being open issues and needs for improvement, the currently available technology already goes a long way to actually enable future system visions to a significant extent. Considering structural challenges, i.e. the question as to how to establish the frame conditions that are required, we think there is no hard blocking point per se. On the other hand, this does not mean this class of challenges is easy or even straight forward to address. The problem here typically lies in questions such as who is responsible for driving what topics, what new structures are required, and so on. Last but not least, assurance related challenges seem to be the most urging ones. The technology capabilities to build systems with high degrees of automation have basically outgrown the established assurance capabilities, so that we momentarily have a mismatch severely hindering progress and innovation. Assurance challenges thus require special attention and additional research, Section 2 sets out to give a more detailed overview in this regard.

On an abstract level, assurance challenges result from unknowns and uncertainties which cannot be resolved during development time due to several reasons. At runtime, however, many of these unknowns and uncertainties could be resolved and resulting implications could then be treated. Thus, a general solution idea is to shift parts of the assurance process into runtime. Figuratively speaking, one could say such approaches are about making systems safety aware much in the way human navigate risk when operating safety-critical systems manually.

Enabling systems in this sense implies equipping them with two key capabilities: 1.) to be aware of the current risk as well as the risk related to intended behaviors, and, 2.) to be aware of own safety related capabilities (i.e., in automotive terms, those related to controllability or reducing severity). The technical means for realizing these capabilities can range from easy checks to sophisticated and intelligent reasoning. It is expected that, in parallel to increasing degrees of adoption of the aforementioned key traits (automation, interconnectedness, AI), systems will need to be empowered to assume increasingly comprehensive runtime assurance responsibilities [10].

Over the last years, runtime safety assurance approaches are increasingly accepted as a promising means by the safety engineering community and corresponding approaches, discussions and considerations can increasingly be found in publications, conferences, working groups and even in standardization groups, such as Digital Dependability Identities [11] being referred to in [12]. At our institute we investigated and developed different building blocks for enabling runtime assurance and devised a superordinate vision called Dynamic Risk Management (DRM). In Section 3 we introduce this vision, a conceptual solution architecture and describe several approaches for the core building blocks of that architecture in more detail.

## 2      Safety Engineering Challenges and Related Work

The trends shown in **Fig. 1** directly translate in specific safety engineering challenges. High levels of automation lead to high levels of system complexity as well as, sometimes even more importantly, context complexity. Interconnectedness and distribution lead to challenges related to communication and interoperability, but particularly also to challenges due to unknowns with respect to the safety-related properties of connected and collaborating systems. Finally, utilization of ML/DL components for e.g. perception tasks inevitably leads to uncertainties, which again constitute a significant challenge for safety assurance. In this Section, we further structure and elaborate corresponding safety engineering challenges.

### 2.1 Assuring Autonomy – How to deal with the complexity?

Increasing automation levels imply increasingly transferring responsibilities from human operators to systems. This naturally translates into increasing complexity of systems, e.g., considering autonomous driving, due to now also assuming tasks related to perception and reasoning. At the same time, the complexity of the physical context of systems takes its toll, as context information is a fundamental input to be provided via perception and reasoned about by the system. Relatedly, with respect to assurance, systems need to be enabled to deal with internal and external risk factors [13]. An elaboration with respect to both of these dimensions follows below.

There is a widely recognized classification of automation levels in the automotive domain standardized in SAE J3016:2021. Automation levels 1 and 2 are purposed to support the driver, whereas the driver is still fully responsible to supervise the system continuously. Starting from level 3, the driver is allowed to take the eyes off the road and to be, at least for a given amount of time, out of the operating loop. However, in level 3, the driver must still be able to step in when the system issues a transition demand with sufficient time to reestablish situational awareness. On paper this seems to be a straightforward fallback mechanism, but the actual practicality is often questioned due to a mismatch between the time required to bring the driver back into the loop and the lookahead capabilities of today's perception stacks [14]. Thus, taking the driver completely out of the loop as foreseen for levels 4 and 5 seems to be the cleaner concept.

Popular definitions of autonomous systems refer to situation-specific behavior. Kagermann et al. [15] define that a system can be described as autonomous if it is capable of independently achieving a predefined goal "in accordance with the demands of the current situation…". A derived requirement is that the system is aware of the risks in the current environmental situation and that it behaves so that the risks will not become unacceptable. In contrast to conventional safety functions, this risk management function requires reasoning about the current environmental situation and potential harm scenarios [16], i.e. the **situational handling of external risk factors**. To this end, safety supervisors are used to monitor the nominal behavior (cf. German application rule VDE-AR-E 2842-6 part 3 clause 12.4), and evaluate the risks of potential system behaviors on different time horizons such as operational, tactical and strategic. Physics-based approaches like Time-To-Collision (TTC) and Responsibility-Sensitive Safety Model [17] are commonly used to formalize risk criteria. However, these physics-based approaches are not sufficient for developing human-like risk assessment capabilities, because they do not take into account knowledge about agent interactions. Thus, expert systems that imitate human risk reasoning must be developed, e.g. with real-time inferrable Bayesian networks [18]. Ultimately, as AI-based perception is often necessary for situational handling of external risk factors due to the required level of situation understanding, the relationship between uncertain AI-based perception and external risk factor estimation needs to be explored. Unsolved challenges in this research area are the identification of relevant risk factors and the definition of a risk criterion combining these factors to express the border between safe and unsafe situations on different time horizons.

Apart from the capability to determine the impact of complex environments on dynamic risk, another important aspect auf autonomous systems is their **self-sufficiency**, autonomicity or capability to maintain operation [19], i.e. achieving a goal without human control [15]. The degree of human independence is promoted in problem structuring frameworks like SAE J3016 or ALFUS [20]. To achieve self-sufficiency, systems need to dynamically handle *internal* risk factors, i.e. all kinds of component failures including failures due to functional insufficiencies or limitations like limited sensor range. **Situational handling of internal risk factors** enhances error detection and handling approaches by considering the risks of detected errors, which vary in different operational situations. Challenges in this research area emerge from the complexity in detecting all relevant errors and handling them so that the best possible functionality is provided. The approach in [21] proposes how to identify component degradation variants and considers the reconfiguration sequences for transitioning from one system configuration to another [22]. Trapp et al. [23] enhances this approach with respect to the risk-oriented handling of errors and proposes to shift parts of a functional safety engineering lifecycle to runtime. Further challenges emerge from components based on Machine Learning (ML), because their output is subject to uncertainty, which we understand as the likelihood that the actual output deviates in a particular way from the correct output. This aspect is further detailed in Section 2.3.

## 2.2  Assuring interconnectedness – How to modularize safety?

Interconnectedness, which enables information sharing, cooperation and collaboration, is a key enabler for future CPS. The potential often seems to be underestimated though. Consider, for example, the field of mobility, where corresponding developments have been going rather slow over the last two decades. Cars (e.g.) are still understood and designed as mostly closed systems, where connectivity is primarily used for infotainment and convenience purposes (and not for optimizing the key functionalities and safety).

Considering safety challenges associated with interconnectedness, safety-related requirements regarding the communication system directly come to mind. But these are often not the most challenging ones, as there usually are ways to forge a sound safety concept around known shortcomings of a communication channel (e.g. black channel concept). A more significant obstacle to overcome are application-level challenges, i.e. assuring the safety of applications/functions that are rendered through a collaboration of different (maybe even dynamically) interconnected systems of different manufacturers. The key issue here is that the safety-related properties of the collaboration partners are not known and that there are no means to determine the actual safety-related guarantees of such a collaboration. Essentially, adequate means for safety modularization are required.

There is a broad range of safety modularization approaches, often contract-based, in the state of the art, mostly focusing on supporting safety engineering for single systems [24,25,26,27,28,29,30,31]. At the same time, there are several approaches taking more of a supply chain perspective, also considering model exchange and tooling, such as the work from the European projects DEIS (the Open Dependability Exchange (ODE) metamodel and its instances called Digital Dependability Identities (DDI) [11]) and AMASS [32].

Utilizing such approaches at runtime was not a topic until relatively recently, John Rushby was one of the first researchers to propose an approach utilizing safety contracts for runtime monitoring to enable "just-in-time certification" [33]. This being still focused on single systems, the next step is to enable CPSoS to dynamically assess their safety-related properties at runtime. The ConSerts approach [34] utilize modular runtime safety models describing safety-related properties for a (constituent) system and a protocol for the dynamic assessment of these properties across composition hierarchies. Further details are provided in Section 3.1. There are further related approaches using runtime models for the assurance of adaptive systems [35][36], distributed service-based systems [37] as well as considerations how a large scale adoption of runtime models might play out to manage heterogenous SoS ecosystems [38] as they are foreseen for the future [10].

Last, it is also noteworthy that safety interfaces and modularization play an important role for shared platforms, e.g. domain controllers, supporting co-existence of different mixed-criticality applications (was e.g. a topic in the European EMC2 project [39]) as well as dynamic updates (e.g. considered in the Up2date project [40]).

## 2.3   Assuring Machine Learning – How to deal with uncertainties?

In addition to interconnectedness, Machine Learning is clearly set to become the second key enabler for future (highly automated) CPS. Perception and behavior planning in autonomous systems are areas where applying ML is attractive and where we have been seeing very good results over recent years. The reason for this is that it is hard to engineer such functions by conventional means, thus it is appealing to let an artificial neural network (NN) learn the wanted behavior based on training data. An example for this is camera-based object detection, where it is difficult to manually specify how to detect e.g. a human. Training an ANN by means of (vast amounts of) training data ideally leads to a strong classification performance with just the right amount of generalization, but realistically this is also no small feat to achieve. Assuring safety is particularly challenging, as there is neither a sound and complete requirements specification nor adequate means to analyze and assure the learned model. There is a range of approaches that can be utilized to attain indications wrt. the performance of an ANN, but there is no silver bullet for generating hard evidence regarding the ANN performing as intended in any situation. As a consequence, assuring systems with ML components typically requires safety concepts which take safety responsibility away from the ML-based channel. **Fig. 2** exemplifies the different elements/channels such a safety concept might encompass, color coding implying a range from "black box" over different shades of grey to "white box". The ML component itself is a black box, but could be transformed into a somewhat greyish box by means of specialized ANN approaches (i.e. assuring the ML component itself), e.g. heatmapping from the field of XAI approaches. Moreover, parallel supervision channels might be established. This might be channels specialized for ANN, e.g. such that the "neuro"-channel is augmented by a "symbolic"-channel. Such approaches are also called hybrid AI approaches and a specific one is briefly introduced in Section 3.3. Apart from that, there is always the possibility to establish AI-agnostic supervisor channels. Those again might include very simple and traditionally realized channels (i.e. easy to assure) that might serve as a last layer of defense, but there also might be more complex elements which again might even include AI or ANN themselves. The idea here would again be that the safety weight is put on the last layer of defense, but the outer layers shall ideally be good enough to ensure that the last layer is never actually needed.

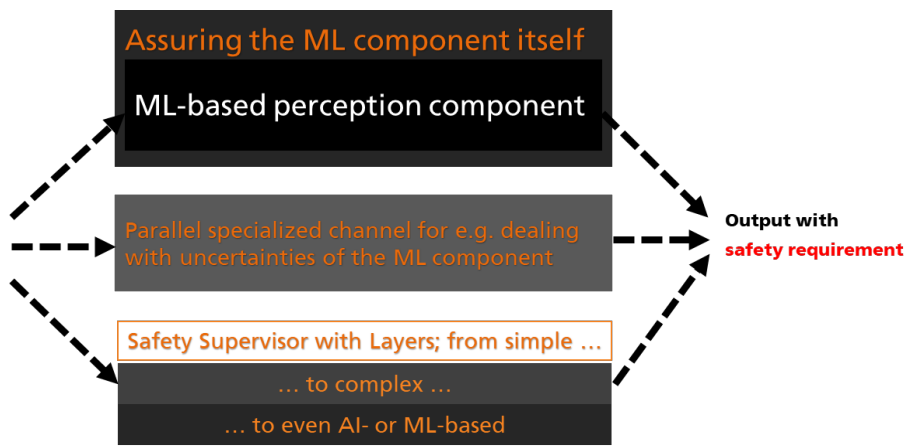

**Fig. 2.** Channels for ML assurance – from "black box" to "white box"

### 2.4  Safety Assurance Frame Conditions

A prerequisite for market introduction of AI systems and autonomous systems is that they are safe. Europe has traditionally very high constraints for market introduction of safety-critical products and is also leading with the introduction of laws for AI systems and autonomous systems. Existing vertical safety regulation is currently enhanced with respect to autonomy. For instance, the new Machinery Directive [41] will replace the Machinery Directive and address autonomous mobile machinery. Regulation for road vehicles, which are not in the scope of the Machinery Directive, was enhanced with respect to automated driving.

In addition to such vertical enhancements of regulation, AI systems are regulated horizontally by the AI Act. The regulatory safety requirements can be quite abstract in order to provide enough stability considering that technology is evolving very fast. Technical safety standards, which can be updated more easily, can provide more concrete requirements. If the safety standards are harmonized, then they can be used to evoke the presumption of conformity to the respective regulations. The development of the harmonized safety standards to the AI Act is currently under development following the standardization request draft [42].

The reason why current safety regulation and standardization is not sufficient is simply because systems were mostly closed systems and had low degree of automation when the standards have been developed. Accordingly, current safety standards provide insufficient guidance for developing cooperative automated systems. In the following, describing our vision of DRM, we illustrate which new topics and questions come up when we focus on high automation level and dynamic cooperation. These topics and questions need be considered when developing novel safety standards.

## 3      Dynamic Risk Management in CPS

Since several years we structure our research on assuring the safety of highly automated CPS according to a scheme we call Dynamic Risk Management (DRM). The main goal of DRM is to improve the performance of automated CPS by means of giving them a better safety awareness. I.e., instead of relying on worst case assumptions, systems shall be aware of their safety-related capabilities and of the current situation to make informed decisions regarding their safe behavior. On an abstract level DRM thus functions similar to how a human operates in critical applications:

 Continuously monitor the current situation. Extrapolate and assess risks, related to both, environment and own behavior intentions. At the same time, be aware of safety-related capabilities of the involved systems. Plan how to operate safely under these circumstances while trying to optimize performance. Implement the plan.

In DRM this process is realized as a MAPE-MART / MAPE-K [43][44] cycle which is illustrated in **Fig. 3**. Monitoring needs to attain the required information for the analysis step in the cycle, i.e. for adequately assessing risk and safety-related capabilities. This comprises 1.) context information for dynamic risk assessment (DRA) via own sensors and/or information from external systems (e.g. edge/infrastructure, cloud), and, 2.) context information for dynamic capability assessment (DCA) by means of internal system monitors as well as information (wrt. their safety-related capabilities) from external systems. Based on this information, the DCA and DRA analyses are conducted, each of which requiring dedicated approaches including runtime models to capture the necessary knowledge (as in MAPE-K). DCA and DRA will be described in more detail in Sections 3.1 and 3.2, respectively. Based on the results of the analysis activities a plan for action is to be devised. We denote this as dynamic risk control (DRC) where the goal is to plan dynamic management of safety-related properties and to optimize the performance of the overall system (while guaranteeing safety).

It is important to note that we are considering a CPSoS scenario, thus there are different collaborating systems from different manufacturers that jointly render complex services or applications. Each of these constituent systems is required to provide (standardized and hence interoperable) MART fragments to adequately support DRA, DCA, DRC and potentially also "safe" sensor-based perception. The latter is mostly about explicitly dealing with perception uncertainties and is detailed in Section 3.3.

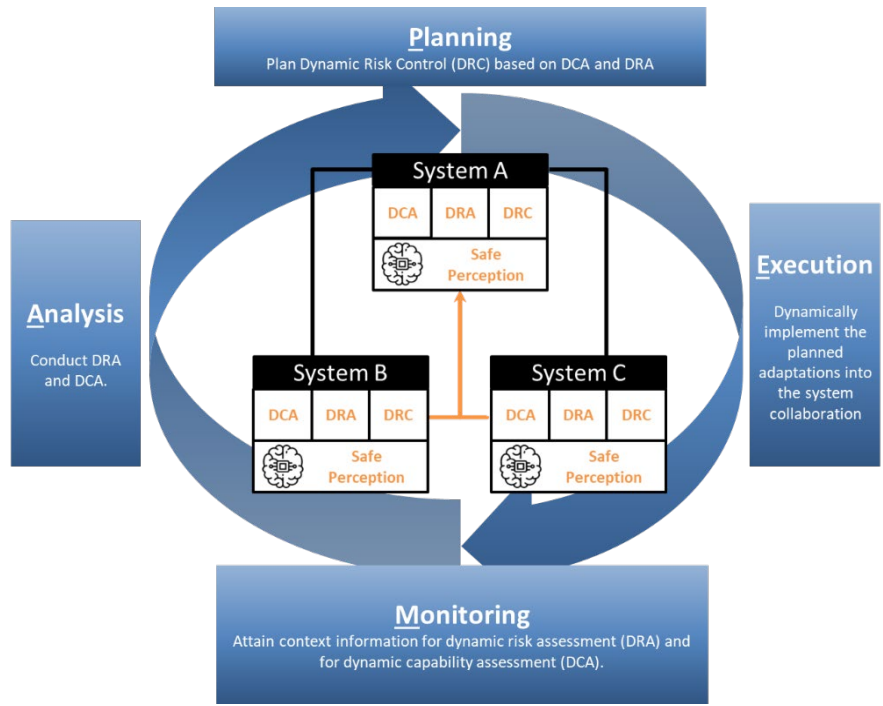

**Fig. 3.** DRM MAPE-MART/K scheme

**Fig. 4** shows a logical runtime architecture for DRM. On the one hand, there are the DRM components (DCA, DRA and DRC) and on the other hand there is the nominal function of the system. In between there is the perception, which provides context information regarding the physical environment for both channels.

It is important to note that using DRM does not mean that anything safety-related is handled by DRM and that the nominal behavior channel is safety agnostic. Rather, DRM provides an additional layer of "safety intelligence", to dynamically adjust safety related parameters (which otherwise would have been fixed at development time) and to continuously optimize performance. As an example, consider that the nominal function channel might implement RSS [17]. RSS is a rule-based safety approach, e.g. considering a safe lateral distance between cars, which is parameterized at development time and rather static (i.e. not context aware) at runtime. As a consequence, RSS might lead to overly cautious driving maneuvers (which does not necessarily improve safety, maybe even to the contrary), because the RSS rules are violated quite often in common human-operated driving [45]. DRM could improve this issue by dynamically adjusting the RSS parameters based on the current DRA and DCA results.

It is further conceivable that parts of the DRM might not be realized with safety integrity. Safety integrity is quantified in current functional safety standards with levels that define which safety measures should be applied to deal with hardware and software-related issues causing incorrect system behavior. For this limited scope, the safety measures are very effective and low target failure rates can be achieved. DRM has a broader scope. By estimating and controlling risks of the current situation, it targets the definition of safe behavior that has to be implemented correctly. For this implementation, it considers AI-based models and it is an open research question if one can find safety measures for such models that are as effective as those for conventional software [69]. DRA in particular might utilize AI-based models, to e.g. realize risk prediction, on top of simpler rule-based layers, thus implementing a layers of protection type of architecture. Ideally, the outer ("low integrity") layers of such an architecture work well enough so that the traditionally forged "last layer of defense" mechanisms are hardly ever required in the field. Still, they of course need to be there to bear the safety responsibility. DCA, on the other hand, can be realized with a high "safety-grade" integrity without too much hassle. ConSerts, for instance, can be developed based on common safety engineering approaches. The only addition being the consideration of variants of assumptions and corresponding guarantees and the integration of rather simple evaluation functions (contracts with variants modeled by means of Boolean logic) into the actual systems.

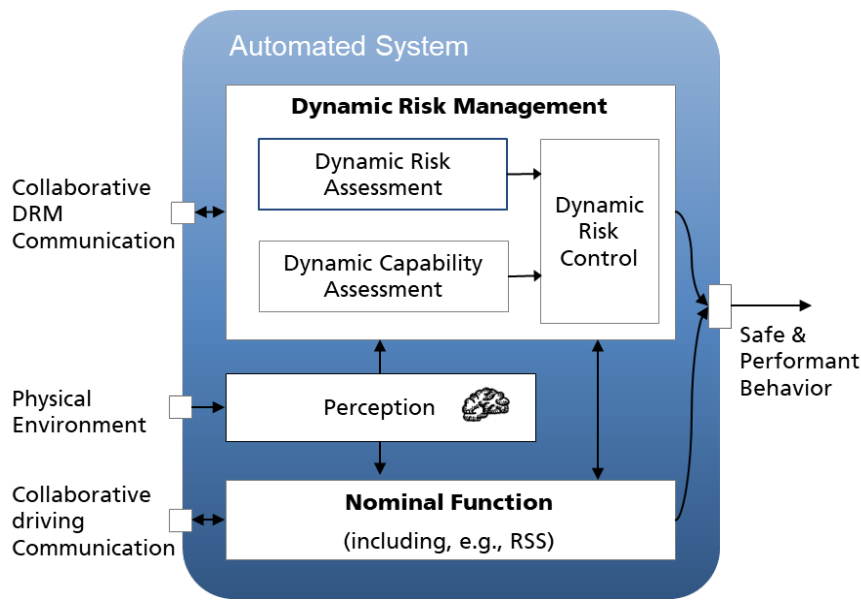

**Fig. 4.** Logical DRM architecture

### 3.1 Dynamic Capability Assessment (DCA)

DCA is about dynamic and continuous assessment of a systems safety-related capabilities. It is the counter-part to dynamic risk assessment, which is introduced in Section 3.2 and which essentially is about a (partially) dynamic hazard and risk analysis, or, in other words, a dynamic (safety) requirements analysis.

On the one hand, a systems capabilities might be dynamically influenced by internal aspects such as system health and resource availability aspects. On the other hand, there are external influences, such as the physical environment or the digital interface to functions/services of other systems. The internal aspects might be directly monitored and their impact on system level functions might be assessed. The external aspects can be monitored by means of corresponding perception capabilities (i.e. aspects that are associated with the physical environment such as weather conditions) and by communication. The latter comprises sharing information between systems (e.g. shared perception) as well as information regarding the safety-related properties of functions involved in a system cooperation. Regarding autonomous systems, one important aspect related external context is the operational design domain (ODD) of the system, which ideally has been defined explicitly and thus can be monitored as well.

Sharing this different kind of information across systems requires corresponding information formats/models and protocols, which today certainly is one significant hindering factor in exploiting the huge potential there is. Without being able to attain sufficient information (with explicit quality guarantees) regarding the environment, worst case assumptions need to be taken, which in turn severely constrain performance.

A key aspect to consider for a DCA approach is how safety responsibility is distributed in a dynamic SoS setting. It is common knowledge that safety is not compositional in a straight forward way, i.e. if you compose several "safe" systems into a SoS, the resulting SoS is not automatically "safe" as well. The issue here is that safety invariably is a property of the overall system in its context and the application/function it realizes. The constituent systems of a SoS cooperation lack this perspective, they can only assume responsibility for rendering their constituent function given that corresponding context assumptions are fulfilled. A solution approach for this aspect is to dynamically form system hierarchies. This implies that cooperative applications do not emerge "by chance", but that there is a system actively orchestrating other so render such an application. This orchestrator would then be, for the duration of that cooperation, be the root of a composition hierarchy and also assume the safety responsibility for the overall cooperation. It would have certain demands wrt. the safety-related properties of the others system functions and it would know what variations in their guarantees would imply regarding the safety of the overall cooperation. Overall, in context of e.g. autonomous driving scenarios, a heterarchical structure would be formed, where e.g. platoons, overtaking maneuvers and so on are rendered based on temporal concurrent application hierarchies.

A DCA approach should provide means for all the above aspects. Conditional Safety Certificates (ConSerts) [34] is such an approach which we have been investigating and applying for over 10 years now. ConSerts operate on the level of safety requirements. They are specified at development time based on a sound and comprehensive safety argumentation (e.g. an assurance case). They conditionally certify that the associated system will provide specific safety guarantees. Conditions are related to the fulfillment of specific demands (i.e. assumptions) regarding the environment. ConSerts comprise different potential guarantees which are bound to different variations of demands. In the same way as "static" certificates, ConSerts shall be issued by safety experts, independent organizations, or authorized bodies (depending on the respective application domain) after a stringent manual check of the safety argument. To this end, it is mandatory to prove all claims regarding the fulfillment of provided safety guarantees by means of suitable evidence and to provide adequate documentation of the overall argument – particularly including the sets of demands and their implications.

Let us briefly illustrate ConSerts based on the example used in [46]. In the agricultural domain, tractor implement management (TIM) enables implements to assume control over the tractor functions, such as setting the vehicle speed or the steering angle. To do this in the best possible way, the implement might consume sensor information from the tractor and from auxiliary third party sensors, such as a swath scanner or a GPS. Consequently, TIM scenarios are scenarios of cooperative CPS, realized by cooperation of different systems of different manufacturers.

For illustrating the engineering of ConSerts in this example the role of the implement manufacturer shall be assumed. The goal of the manufacturer is to develop a round baler with TIM support. From a functional point of view, it is clear (due to existing AEF [47] guidelines) how the interfaces between the potential participants look like and how they are to be used. However, the implement manufacturer does not know about the safety properties of these functions.

From a safety point of view, the engineering of the baling application starts top-down with an application-level hazard and risk analysis. Assume that the agricultural manufacturers agreed by convention that during the operation of a TIM application, the application (and thus the application manufacturer) has the responsibility for the overall cooperation, thus dynamically forming a cooperation hierarchy as motivated above. Therefore, the safety engineering goal is to ensure adequate safety not only for the TIM baling application or for the implement, but for the whole cooperation of systems that will be rendering the cooperative application at runtime. Application-level hazards of the TIM baling application could correspondingly comprise the tractor having an unwanted acceleration or steering during TIM baling. Causes might be located in the TIM baling application itself or in the tractor or in other cooperating systems (e.g. a third-party sensor). Causes in the TIM baling application and the implement are tackled by traditional safety engineering. Causes outside the system under development are translated into ConSert demands and runtime evidences that are to be evaluated at runtime. Thanks to the ConSert-based modularization it is thereby sufficient to only consider the direct dependencies of the system under development on its environment. The runtime evaluation can be done bottom up, i.e. the system at the leaves of the cooperation hierarchy determine their guarantees and propagate them up so that the root (here: the TIM baling application) can determine its guarantees. Based on these guarantees, the cooperation might be parameterized (e.g. set maximum speed) to ensure safety but also maximize performance.

Enabling dynamic management of system performance while always ensuring safety is a key benefit of DCA in general. From changes in system guarantees due to wear and tear to changing weather conditions, anything can potentially be considered. Thus, it is no longer necessary to work based on worst case assumptions (because you cannot know the actual conditions during operation), with DCA systems become aware regarding the safety-relevant context conditions.

Of course, these benefits come at a certain price because additional engineering is required, where some technical aspects might also be automated as investigated in [48]. Overall, ConSerts are a relatively lightweight DCA approach and they are well aligned with traditional safety engineering. The main difference being that unknown context is structured into a series of foreseen variants, which are then specified in terms of a relatively simple modular runtime model. While this already provides significant gains in terms of flexibility and realizable system performance (compared to a completely static approach), there is still further potential. More powerful DCA evaluation logic could be investigated beyond simple contracts with variants, e.g. with the capability to calculate with probabilities and with fuzziness.

### 3.2 Dynamic Risk Assessment (DRA)

DCA monitors aspects affecting the CPS capabilities to react dependably in a critical situation. However, whether a particular error or failure is safety-critical and poses an actual risk depends on the current operational situation the system finds itself in during runtime. Hazardous events and their associated risk, i.e. the likelihood of a transition from a system hazard to an accident as well as the potential severity of this accident, are always conditioned on the operational situation. Since the operational environment in the envisioned CPS use cases is highly dynamic, determining the impact of the current situation on the risk parameters at runtime can increase system performance, as the CPS may continue operation in situations, where it would have stopped in conventional worst-case assumed environmental situations. Consequentially, if we monitor the presence of risky/non-risky operational situations, we can increase CPS performance by a) being able to tolerate certain faults and failures if their associated risk is low in the current situation and b) actively reducing particular risk parameters with tactical decisions, where severity, controllability or the operational situation itself is changed. Dynamic risk assessment (DRA) techniques thus treat the CPS or a constituent system as a black box and provide means to analyze the consequences of CPS behavior deviations or present system hazards on the risk in the current operational situation.
For autonomous systems, in particular in the automotive domain, DRA techniques have been applied to address the trade-off between performance and safety risks. The existing approaches can be classified broadly into three categories. 1. DRA is incorporated into motion and trajectory prediction frameworks by specifying behavioral or kinematic constraints that are input to the trajectory planner [49,50]. 2. DRA is performed during the online verification of an already planned yet potentially unsafe trajectory [51,17]. 3. DRA is not performed in direct relation to a planned trajectory, but instead monitors risk parameters that enable distinction between the presence/absence of a hazardous event and is used to adapt the parameters of risk metrics used in trajectory planners [52, 53, 54].
Although different architectures for incorporating DRA into autonomous systems exist, all of them need to decide which particular risk-influencing situation features are to be observed in the present, project the evolution of these features into the future by using a set of assumptions, and rate the risk of the projected future situation. For each of these DRA sub-tasks, respective design time engineering activities are required. Detailed information on situation prediction techniques and concrete risk metrics can be found in relevant literature reviews (see [55], [56], [57]). To enable machines to perform DRA, modelling formalisms are required to technically capture the relationship between situation features and risks. For this purpose, different classes of models can be used, which are model-based (e.g. structural equation models), rule-based (Boolean models), generative (Gaussian Mixture Models, dynamic Bayesian networks, Hidden Markov Models), discriminative (Decision Tree, Random Forest, Support Vector Machines) or deep learning-based (Multi-Layer Perceptron, Convolution Neural Network – CNN, Recurrent Neural Network – RNN, Long-Short-Term-Memory Network – LSTM). Some of them are developed purely based on expert knowledge, and some of them can be adapted to new operational domains with machine learning techniques. Depending on the concrete modelling formalism selected, support for deterministic or probabilistic assumptions may be given as well as the consideration of uncertainties during feature perception is possible.

At Fraunhofer IESE, [55] introduced a concrete DRA approach for autonomous systems. The work contributed a conceptual taxonomy of DRA and provided a concrete application using concrete instances of controllability and severity metrics to rate collision risks for automated driving. In order to be applicable in many different operational situations, risk metrics were required to be situation-agnostic. Situation-specific features such as road structure, lighting and weather conditions, traffic rules, and actor interactions were consequentially not considered. To account for the fact that risk is influenced by exactly those situation-specific features, the work in [55] was extended to the Situation-Aware Dynamic Risk Assessment (SINADRA) framework [58]. The approach uses Bayesian networks as a modelling formalism and explicitly considers the mentioned situation-specific features as risk influences. A proof-of-concept and tool implementation of the approach is presented in [59]. The design-time method for building Bayesian situation prediction models with machine learning techniques is presented in [60]. More recently, SINADRA runtime models have been formally traced to model-based design-time safety engineering artifacts and applied in a distributed smart logistics mobility system [61]. An important aspect of a DRA model is its connection to a model-based representation of the operational design domain (ODD). First steps in this direction have been explored in [62].

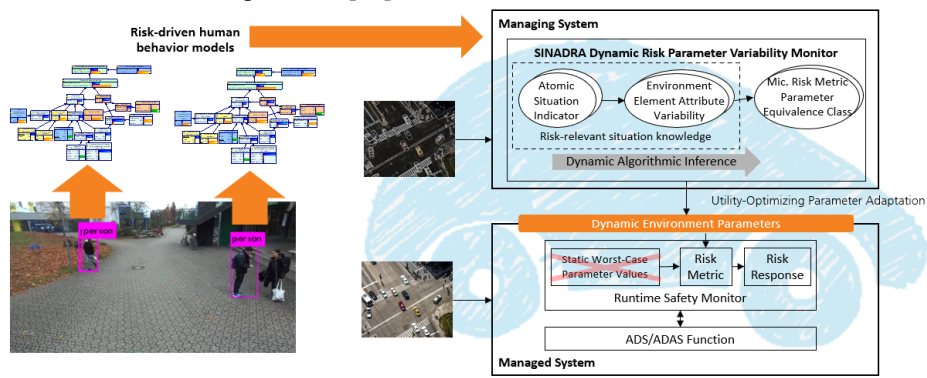

**Fig. 5.** Logical DRA architecture

The conceptual components of the SINADRA framework are shown in Figure 5. The core artefact of the framework is the DRA Monitor Runtime Component. It is an executable software component that is deployed to a CPS along with the nominal functionality. Functionally, it contains a model that captures the variability of risk parameters with respect to risk-relevant atomic situation indicators. Thus, the output is a dynamic risk classification of particular (unsafe) behavior intentions of interest. Based on the runtime perception of relevant situation features and the inference determining the effect of the dynamic input feature vector on risk, the basis for a risk-informed parameter adaptation is given.

### 3.3   ML-powered Perception

Adequate perception capabilities are a key ingredient for safe autonomy and also important inputs for DCA and DRA. As pointed out in Section 2.3, ML/DL based perception is in the limelight and there presently is a lot of attention being spend in research to get a grip wrt. the uncertainties associated with the output of ML/DL based components.

The idea is to measure uncertainty during operation and to use the explicit uncertainty information in DCA and DRA. A consequence of a high uncertainty value could thus be that the current sensor capabilities are assessed to be low (DCA) and that this propagates through to an ongoing cooperation scenario. As a consequence, some parameters of the cooperation (e.g. min distance in platooning) are adjusted to ensure safety. Likewise, the current risk might be assessed conservatively (i.e. as being high) due to perception uncertainties (DRA) and again this leads to adaptions to maintain safety. At this point, there might be an overlapping between DCA and DRA which needs to be partitioned carefully.

The general scheme has some likeness to the "classical" error detection and error handling. Considering traditional software errors, the error will always occur under some specific conditions. This holds also true for software that is developed by means of ML/DL, but the conditions can be hard to grasp in this case. Consider e.g. the well-known examples of adversarial attacks on traffic sign detection [63]. Still, general environmental conditions can be considered such as noise, e.g. weather/fog for a camera-based object classification. Depending on such conditions there is a likelihood that the current output of an ANN deviates in a particular way from the correct output. One approach to estimate this likelihood during operation is given by an uncertainty wrapper (cf. DIN SPEC 92005 and [64]). In many cases this uncertainty is quite high so that remediating action would be triggered too often. An approach to deal with this issue is to aggregate uncertainties from different time steps [65]. Additionally, simple architectural patterns to handle uncertainty by increased safety margins are proposed in [66]. A complementary approach which we are investigating is based on statistical similarity metrics applied to training data on the one hand and to field/test data on the other [67, 68]. The general rationale here being that if the system encounters input data which is very dissimilar to the training data, there is a high level of uncertainty regarding the correctness of the output of the ML component.

Overall, more research is still required to facilitate safety assurance of systems with ML components. There are numerous approaches that can provide indications about potential flaws of a learned model (e.g. approaches from the field of explainable AI), but there is a lack of means to attain strong evidence that a ML component will always function as intended. Thus, it is necessary to work around this issue and assign safety responsibility to other redundant channels as illustrated in **Fig. 2**.

## 4      Summary and Conclusion

In this article we described key trends of future CPSoS, explained corresponding safety assurance challenges and summarized some of the existing work to tackle these challenges. Then we went on to introduce dynamic risk management, our vision for making future systems safety-aware. We introduced a conceptual architecture for DRM and presented some of our work regarding the key building blocks of that architecture.

DRM is not to be understood as a concrete one-size-fits- all solution, but rather as a vision and a framework for structuring various approaches into a comprehensive perspective. Consequently, DRM could be instantiated for many different application domains and applications scenarios, maybe using some of the approaches mentioned here, or others that are deemed suitable.

DRM touches all the challenges identified in Section 2. Overall, DRM is aimed at safety assurance and performance optimization in automated and autonomous systems. In this context, DRA is specifically dealing with the physical context complexity, DCA with safety modularization (which is also a means for dealing with system complexity) and interconnectedness, and perception has to deal with the uncertainties associated with ML components.

Each of the presented dimensions of challenges still requires additional research, whereas the assurance of ML components might be the one with the highest need. But also on the integration level, where DRM as a whole is located, we are still just at the beginning. Safety awareness and resilience are key characteristics regarding the competitiveness of future CPSoS products, because these not only need to be safe, but they also need to be available and performant. Indeed, there clearly is the potential for future systems to be better and safer than today's systems by a huge margin. Given the significant transformations the industries in the embedded domain are undergoing due to the described trends, those who have the best answers to these questions will be significantly strengthening their position in the global competition.

Looking even further into future as illustrated by **Fig. 1** and, for instance, described in the safeTRANS roadmap "Safety, Security, and Certifiability of Future Man-Machine Systems." [10], in the long run systems are set to assume ever higher degrees of autonomy, interconnection and utilization of AI – requiring further research with respect to safety assurance and certifiability, significantly extending what has been presented here.

17# References